# Annealing-dependence of Exchange Bias in MnO/$Ga_{1-x}Mn_xAs$ Heterostructures

K. F. Eid, O. Maksimov, M. B. Stone, P. Schiffer, N. Samarth[1]

*Department of Physics and Materials Research Institute, Pennsylvania State University,*

*University Park, PA 16802*

We discuss the annealing-dependence of exchange bias in $Ga_{1-x}Mn_xAs$ epilayers that are overgrown with Mn. Although pure Mn is a known antiferromagnet, we find that the as-grown Mn overlayer does not produce any exchange coupling with $Ga_{1-x}Mn_xAs$. Rather, our studies indicate that annealing in air is essential for creating the standard signatures of an exchange bias (a clear shift in the magnetization hysteresis loop and an increased coercive field). We use x-ray photoelectron spectroscopy to characterize the compositional depth profile of both as-grown and rapid thermal annealed samples. These studies demonstrate that the cleanest exchange bias arises when the Mn overlayer is completely converted into MnO.

**KEY WORDS**: ferromagnetic semiconductor, exchange bias, (Ga,Mn)As

[1] Corresponding Author: nsamarth@psu.edu

# 1. Introduction

Ferromagnetic semiconductors derived from III-V semiconductors such as GaAs and InAs have attracted significant attention recently within the context of semiconductor-based "spintronics" [1,2]. These materials present a unique opportunity for controlling magnetism through electrical gates [3], and have also been incorporated a variety of proof-of-principle device configurations [4,5,6]. The ability to exchange bias spintronic devices derived from these materials will add to their viability for future applications, as well as offer a new system for basic scientific research. For example, it is known from studies of metallic ferromagnetic multilayer structures that exchange biased read heads are more stable than hybrid sensors [7]. Exchange biased spin-valves are also preferred for devices because of the desirable shape of their magneto resistance curves [8].

Exchange biasing of a ferromagnet by a vicinal antiferromagnet (AF) manifests itself as a shift of the magnetization hysteresis loop, making it centered about a nonzero magnetic field (the exchange field, $H_E$). This occurs when the sample is cooled through the blocking temperature ($T_B$) of the AF in the presence of either an external magnetic field or a net magnetization in the ferromagnetic layer [9]. This shift is typically accompanied by an increase in the coercive field ($H_C$) of the ferromagnetic layer. This phenomenon is the subject of intense study in a variety of magnetic materials and is exploited in different applications [9].

We have recently demonstrated the first successful exchange biasing of a ferromagnetic semiconductor ($Ga_{1-x}Mn_xAs$) by an overgrown antiferromagnet (MnO) [10]. In these experiments, metallic Mn is first epitaxially grown on top of $Ga_{1-x}Mn_xAs$ in an ultrahigh vacuum chamber. Upon removal from the vacuum chamber, the Mn overlayer is unintentionally oxidized during removal of the wafer from the indium-bonded mounting block. Rutherford backscattering (RBS) measurements clearly indicate the presence of MnO, a well-known antiferromagnet that is presumably responsible for the exchange biasing effect. This conclusion is supported by the observation that Al-capped samples do not exhibit any exchange bias since they are impervious to oxidation. Furthermore, RBS studies also suggest that the extreme reactivity between Mn and GaAs requires a delicate balance between the need for oxidation and the preservation of a clean interface. For instance, modest annealing at temperatures around 230 $^o$C completely destroys the exchange bias.

Given these circumstances, it is important to develop a better understanding of the oxidation of the Mn overlayer and to hence obtain systematic control over the creation of exchange biased $Ga_{1-x}Mn_xAs$ devices. In this paper, we focus on the growth and characterization of the antiferromagnetic MnO layer and discuss the dependence and sensitivity of exchange coupling on the annealing conditions. In particular, we use a series of annealing experiments to definitively demonstrate that metallic Mn grown on $Ga_{1-x}Mn_xAs$ does not by itself create exchange bias, and that the oxidation of the Mn layer to MnO is essential to produce an exchange coupling.

## 2. Experimental Details

Samples are grown by molecular beam epitaxy (MBE) on "epi-ready" semi-insulating GaAs (100) substrates that are bonded with indium to molybdenum blocks.

Removal of the samples from the mounting block entails heating up the entire block to melt the bonding layer of indium. In contrast to the usual mounting, when indium is uniformly dispersed under the entire wafer, for samples studied in this communication only the edges of the sample wafer are indium-bonded to the block. This procedure provides sufficient thermal contact during MBE growth and allows the removal of a significant portion of the wafer without any post-growth annealing.

The MBE growth is performed in an Applied EPI 930 system equipped with Ga, Mn, and As effusion cells. The substrates are deoxidized using standard protocol, by heating to ~ 580 $^0$C with an As flux impinging on the surface. A 100 nm thick GaAs buffer layer is grown after the deoxidization. Then, samples are cooled to ~ 250 $^0$C for the growth of a 5 nm thick low temperature GaAs layer, followed by a 10 nm thick $Ga_{1-x}Mn_xAs$ layer ($x \sim 0.06$). The growth is performed under the group V rich conditions with a As:Ga beam equivalent pressure ratio of ~ 12:1. After the $Ga_{1-x}Mn_xAs$ growth, the samples are transferred *in situ* to an adjoining ultra high vacuum buffer chamber and the As cell is cooled to the rest temperature of 110 $^0$C to avoid formation of MnAs clusters during the Mn growth. When the As pressure in the growth chamber decreases to an acceptable level, the wafers are reintroduced. Then, the Mn capping layer with a thickness of ~4 nm or ~8 nm is deposited. Mn growth is performed at room temperature in order to prevent inter diffusion and chemical reaction between Mn and $Ga_{1-x}Mn_xAs$ layers [11,12].

The growth mode and surface reconstruction are monitored *in situ* by reflection high-energy electron diffraction (RHEED) at 12 keV. The thickness of the $Ga_{1-x}Mn_xAs$ layer is calculated from RHEED oscillations, while the thickness of the Mn layer is estimated from RHEED oscillations of MnAs whose growth rate is mainly determined by the sticking coefficient of Mn. The Mn concentration in $Ga_{1-x}Mn_xAs$ is $x \sim 0.06$, estimated from the Mn flux.

Magnetization measurements are performed using a commercial superconducting quantum interference device (SQUID) magnetometer. Samples were measured with the magnetic field in plane along the (110) direction as a function of both temperature and applied magnetic field. Surface morphology is investigated using an atomic force microscope running in a tapping mode. Si cantilevers with a spring constant of 13 – 100 N/m and a nominal resonance frequency of 240 – 420 KHz are used. Images are recorded with a resolution of 300 x 300 pixels at a scan rate of 1 μm/s. The surface and sub-surface composition is examined by x-ray photoelectron spectroscopy (XPS) using a Kratos Analytical Axis Ultra system. The Ga $2p_{3/2}$, As $2p_{3/2}$, Mn $2p_{3/2}$, $2p_{1/2}$, 3p, and O 1s photoelectrons are excited using a monochromatic Al Kα x-rays (hv = 1486.6 eV). XPS quantification is performed by applying the appropriate relative sensitivity factors (RSFs) for the Kratos instrument to the integrated peak areas. These RSFs take into consideration the x-ray cross-section and the transmission function of the spectrometer. Low energy (<5 eV) electrons are used for charge neutralization. The charge-induced shifts are corrected using the binding energy of C-C hydrocarbon line (285.0 eV) as an internal standard.

## 3. Results and Discussion

Figure 1 (a) shows the (1x2) surface reconstruction in a RHEED pattern for a $Ga_{1-x}Mn_xAs$ epilayer prior to the Mn growth. Figures 1 (b) and (c) show RHEED patterns for Mn deposited on $Ga_{1-x}Mn_xAs$ with the incident electron beam along the [110] and [1$\bar{1}$0] directions. The RHEED pattern consists of sharp, elongated streaks and its symmetry is suggestive of the stabilization of a cubic phase of Mn. This indicates that, in contrast to previous reports [11], Mn can be epitaxially grown at room temperature. A representative atomic force microscope scan of the Mn surface is shown in Figure 2. The epilayer is atomically flat with a root mean square (rms) roughness of ~ 0.2 nm, similar to the rms values reported for $Ga_{1-x}Mn_xAs$ epilayers [13].

Figure 3(a) shows the temperature dependence of the magnetization for a sample with a Curie temperature ($T_C$) of 85 K. The data is shown for two pieces from the same wafer: one piece is from the indium-free portion of the wafer and is not heated after removal from the vacuum chamber, while the other is from the indium-bonded portion and hence undergoes a rapid thermal anneal to ~ $220^0$ C during sample removal. The zero magnetization at temperatures above $T_C$ indicates that the sample is of good quality without MnAs clusters. Although we observe no difference in the $T_c$ of the indium-free and indium-mounted portions of the sample, we do note that the former has a smaller low-temperature saturated moment compared to the latter. Figures 3(b)-(c) show the magnetization (*M*) of the bilayer structure as a function of the applied magnetic field (*H*) after the samples are cooled to the measuring temperature (*T* = 10 K) in the presence of an external magnetic field of 1 kOe. Figure 3(b) shows the magnetization of an indium-free part of the wafer. The magnetization curve is symmetric about zero applied field, indicating absence of exchange bias. Figure 3(c) shows a shifted hysteresis loop measured for an indium-mounted portion of the sample. Finally, Figure 3(d) shows a hysteresis loop of an indium-free portion of the sample that was intentionally annealed at 200⁰ C for 1 min. The center of the hysteresis loop is also shifted from zero. These results clearly demonstrate that annealing is necessary to create exchange bias in the Mn/Ga$_{1-x}$Mn$_x$As bilayer structures.

To further understand these results, we perform detailed XPS studies on the indium-free portion of the wafer. We study the as-grown sample, as well as pieces that are annealed for 1 minute at 200 $^0$C and 250 $^0$C. Since qualitatively similar spectra were collected for the samples annealed to 200 $^0$C and 250 $^0$C, we only compare the 250 $^0$C annealed sample to the as-grown sample. Figure 4 depicts high-resolution Mn 2p spectra for the as-grown and annealed samples. Both the positions and the shapes of the Mn 2p lines differ for the annealed and as grown samples. The Mn 2p$_{3/2}$ line for the as grown part is centered at ~ 641.0 eV. Its position is in agreement with the binding energy of MnO (Mn$^{2+}$), indicating that the Mn at the surface is oxidized. (Metallic Mn$^0$ has a 2p$_{3/2}$ line at 639 eV, while lines from Mn$_2$O$_3$ (Mn$^{3+}$) and MnO$_2$ (Mn$^{4+}$) have binding energies of ~ 641.7 eV and ~ 642.5 eV, respectively [14]). The observation of two satellite lines spaced by 5.5 eV from 2p$_{3/2}$ and 2p$_{1/2}$ lines further supports this assignment. These satellite excitations are typical for MnO [15, 16] and are not present in either Mn$_2$O$_3$ or MnO$_2$ [17,18,19,20]. The Mn 2p$_{3/2}$ line for the annealed piece is visibly shifted to a slightly higher binding energy of ~ 641.3 eV. The two satellite lines disappear while a new satellite line at ~ 663.5 eV appears, indicating that the as-grown Mn has oxidized to Mn$_2$O$_3$ [17].

We also measure the compositional depth dependence in ion milled samples using 4 keV Ar$^+$. Figure 5(a) shows the relative percentage-depth profile of Mn, O, Ga, and As for the as-grown part. Even prior to sputtering, As (Ga) 2p lines are detected. This does not necessarily indicate As (Ga) diffusion and is, most probably, due to the fact that the Mn film thickness is comparable to the probing depth. The data show that the As (Ga) content increases with depth, while the Mn content decreases. The change is linear and the content becomes constant after 150 s, indicating the depth of the $Ga_{1-x}Mn_xAs$ interface. A Significant amount of O is present close to the surface with an O/Mn ratio of approximately 1. However, O content drops rapidly with depth, and only traces are present after 100 s with the O/Mn ratio consistently less than 1/3. This suggests that only the top portion of the film is oxidized and metallic $Mn^0$ is present closer to the $Ga_{1-x}Mn_xAs$ interface.

Figure 5(b) illustrates the relative percentage-depth profile for the annealed piece of the indium-free mounted sample. In contrast to the as-grown part, we do not detect As (Ga) lines prior to sputtering, and the As (Ga) traces become evident only after sputtering for 100 s. This is followed by a linear change in content, and finally a stabilization in concentration after 250 s. A larger O content is present with O being more uniformly distributed with an O/Mn ratio of ~1. Thus, the Mn film is more significantly oxidized upon annealing. We furthermore conclude that oxidation results in the expansion of the crystal lattice (~ 1.8 times assuming complete conversion of Mn° to MnO). Since the MnO film is thicker, we do not see As (Ga) lines when the top layers are studied.

This interpretation is further supported if we consider Mn 2p lines as a function of depth (Figure 6). The Mn $2p_{3/2}$ line from the annealed sample is characteristic for MnO. Its position and shape remains constant and only the intensity decreases due to a decrease in Mn content. Thus, the annealed film is nearly uniformly oxidized with MnO being the dominant form of Mn throughout the whole layer. In contrast, the Mn $2p_{3/2}$ line from the as-grown part exhibits a low energy shoulder after 60 s and shifts to 639 eV after 90 sec. Satellite lines also disappear at this point. This clearly indicates that while the surface layers of the as-grown part are oxidized, metallic $Mn^0$ dominates in the bottom layers.

## 4. Conclusions

In summary, we have demonstrated that epitaxially grown $Mn/Ga_{1-x}Mn_xAs$ heterostructures do not any exhibit exchange bias until the Mn layer has been completely oxidized into MnO by thermal annealing in air. XPS measurements as a function of depth show that the as-grown Mn layer oxidizes partially into MnO upon exposure to air, with some metallic Mn regions close to the $Mn/Ga_{1-x}Mn_xAs$ interface, as well as some regions of $Mn_2O_3$ at the surface. Rapid thermal annealing to ~200 $^o$C completely converts the Mn layer into MnO, creating an exchange-biased system.

**Acknowledgements**

The research has been supported by DARPA/ONR under grants N00014-99-1093, and -00-1-0951, by ONR N00014-99-1-0071 and by NSF DMR 01-01318. We thank J. Shallenberger for useful discussions and assistance with XPS measurements.

**Figure Captions:**

Fig. 1. (a) RHEED patterns for $Ga_{1-x}Mn_xAs$ showing the 2-fold reconstruction prior to Mn growth; (b) and (c) RHEED patterns for Mn epilayers grown on $Ga_{1-x}Mn_xAs$ with different incident directions of the electron beam.

Fig. 2. Tapping-mode atomic force microscope scan of an as-grown $Mn/Ga_{1-x}Mn_xAs$ heterostructure with 8 nm Mn thickness.

Fig. 3. Magnetization as a function of temperature and applied magnetic field for $MnO/Ga_{1-x}Mn_xAs$ heterostructures. Hysteresis loops are measured at $T = 10$ K, field cooled from $T = 200$ K with $H = 1$ kOe. (a) Temperature-dependent magnetization measured with $H = 100$ Oe for an indium-free portion of the as-grown sample and for an indium-bonded portion. (b) Field-cooled hysteresis loop for indium-free portion of sample. (c) Field-cooled hysteresis loop for indium-mounted portion of sample. (d) Field-cooled hysteresis loop for indium-free portion of sample after intentional annealing at $T = 200\ ^\circ C$ for 1 minute. A diamagnetic background has been subtracted from the data shown. A paramagnetic background has been subtracted from the data shown in panels (a)-(c).

Fig. 4. The Mn 2p spectra acquired at the surface for the (a) annealed and (b) as-grown In-free parts of the $Mn/Ga_{1-x}Mn_xAs$ heterostructure.

Fig. 5. Relative concentration as a function of depth of Mn, O, As, and Ga in the (a) as-grown and (b) annealed indium-free parts of a $Mn/Ga_{1-x}Mn_xAs$ heterostructure. The nominal thickness of Mn deposition is 8 nm.

Fig. 6. The Mn 2p spectra acquired at the different depths for (a) annealed and (b) as-grown indium-free piece of a $Mn/Ga_{1-x}Mn_xAs$ heterostructure.

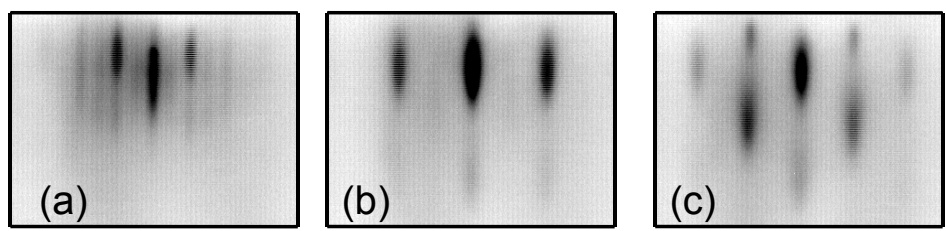

Figure 1

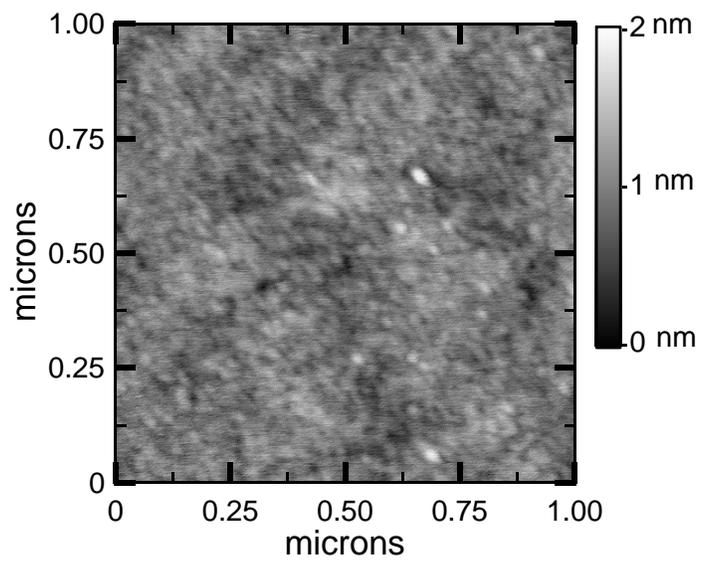

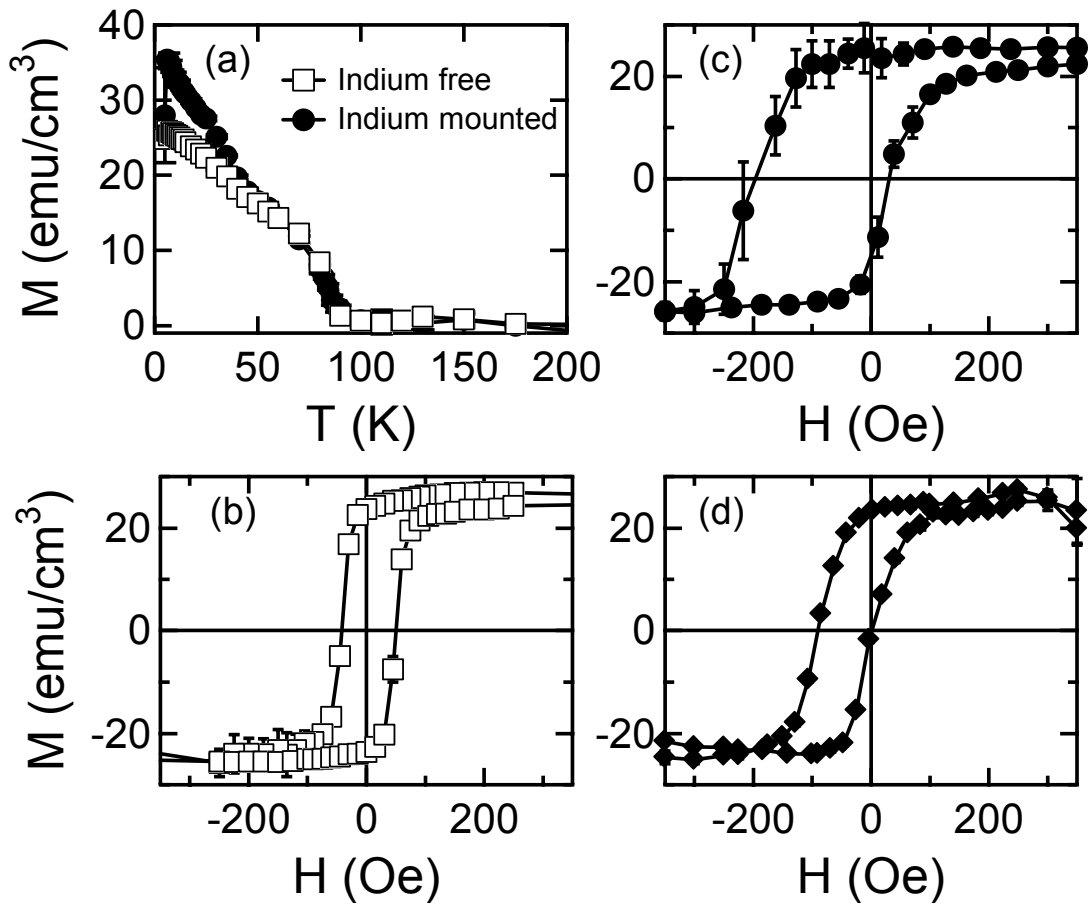

Figure 3

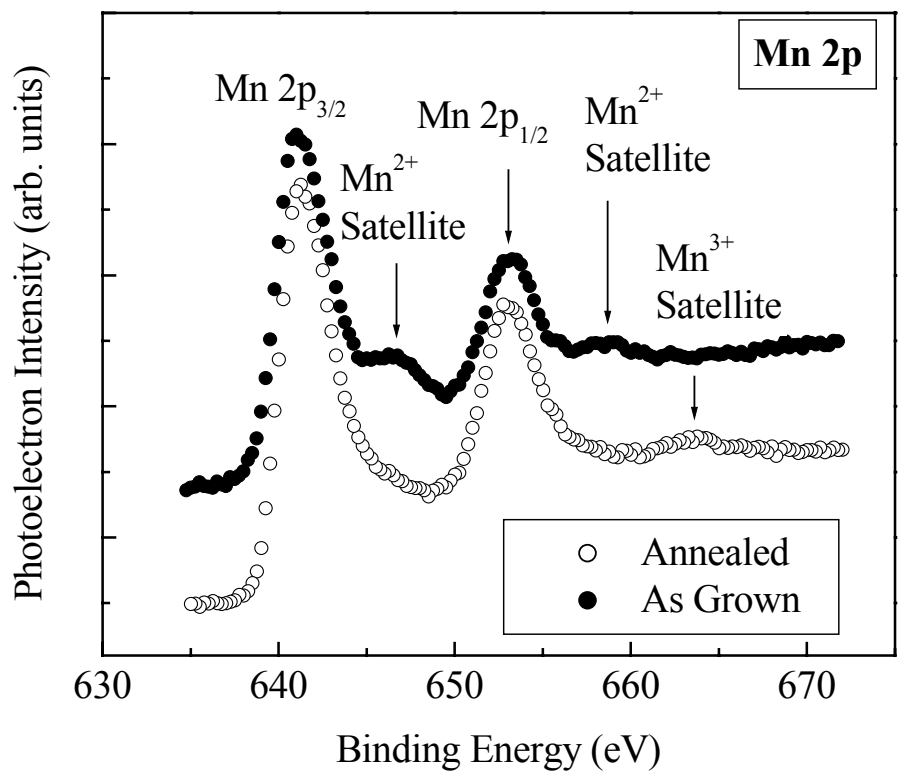

Figure 4

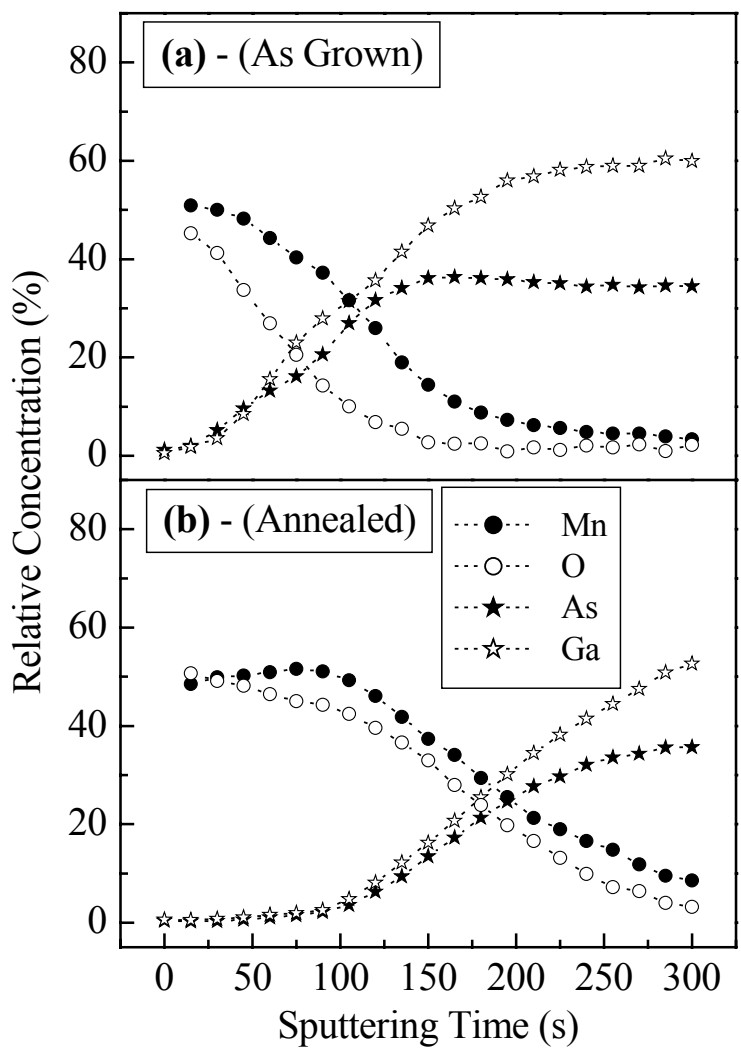

Figure 5

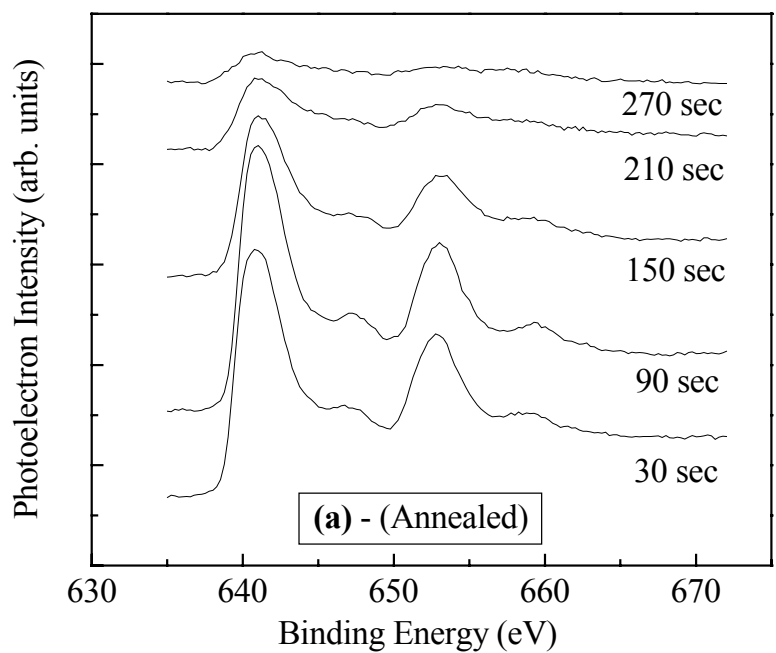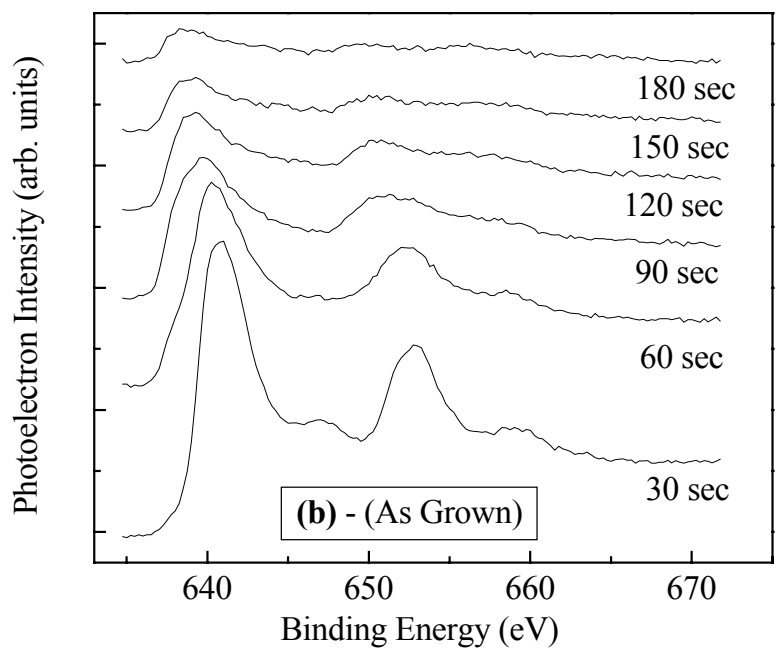

Figure 6